\documentclass{SciPost}

\hypersetup{
    colorlinks,
    linkcolor={red!50!black},
    citecolor={blue!50!black},
    urlcolor={blue!80!black}
}

\usepackage[bitstream-charter]{mathdesign}
\DeclareSymbolFont{usualmathcal}{OMS}{cmsy}{m}{n}
\DeclareSymbolFontAlphabet{\mathcal}{usualmathcal}

\fancypagestyle{SPstyle}{
\fancyhf{}
\lhead{\colorbox{scipostblue}{\bf \color{white} ~SciPost Physics }}
\rhead{{\bf \color{scipostdeepblue} ~Submission }}

\fancyfoot[C]{\textbf{\thepage}}
}

\usepackage{changepage} 

\begin{document}

\pagestyle{SPstyle}



\newcommand{\DWBA}{DWBA}
\newcommand{\SRID}{SRID}
\newcommand{\ME}{ME}
\newcommand{\RE}{RE}
\newcommand{\SE}{SE}

\newcommand{\EWDLS}{EWDLS}
\newcommand{\RI}{RI}
\newcommand{\RP}{RP}
\newcommand{\EpsNull}{\varepsilon_0}
\newcommand{\MuNull}{\mu_0}
\newcommand{\EpsRel}{\varepsilon_{\mbox{\scriptsize r}}}
\newcommand{\Muriel}{\mu_{\mbox{\scriptsize r}}}
\newcommand{\ETMT}{ETMA}
\newcommand{\GISAXS}{GISAXS}
\newcommand{\FAC}{FAC}
\newcommand{\DRY}{DRY}
\newcommand{\WET}{WET}
\newcommand{\DD}{DD}
\newcommand{\MAME}{MAME}
\newcommand{\MIME}{MIME}
\newcommand{\CRs}{CRs}
\newcommand{\RT}{RT}
\newcommand{\CC}{CC}
\newcommand{\IBF}{IBF}
\newcommand{\PCCP}{PCCP}
\newcommand{\AI}{AI}
\newcommand{\ETMA}{ETMA}
\newcommand{\JFree}{\mathbf{J}^{\mbox{\scriptsize (free)}}}
\newcommand{\RhoFree}{\rho^{\mbox{\scriptsize (free)}}}
\newcommand{\ASAXS}{ASAXS}

\begin{center}{\Large \textbf{\color{scipostdeepblue}{
Domain of Applicability of the\\
Distorted Wave Born Approximation (\DWBA):\\
A Critical Analysis for X-ray and Visible-Light Interface Scattering\\
}}}\end{center}


\begin{center}
R. Sigel\textsuperscript{1$\star$}
\end{center}

\begin{center}
{\bf 1} Independent Scientist, 88677 Markdorf, Germany
* res314159@aol.de
\end{center}



\section*{\color{scipostdeepblue}{Abstract}}
\textbf{\boldmath{%
%
The Distorted Wave Born Approximation (\DWBA) is established for X-ray interface scattering, yet its domain of applicability has never been stated explicitly. The \DWBA\ implicitly rests on the small refractive index deviation (\SRID) condition — satisfied by X-rays ($|n-1| \sim 10^{-6}$) but violated by visible light ($|n-1| \sim 0.5$). The \DWBA\ mixes microscopic and macroscopic Maxwell equations, whereas a rigorous optical theory must be built on the macroscopic Maxwell equations and constitutive relations. Rigorous alternatives for two specific inconsistencies in the standard \DWBA\ are discussed. The Extended Transfer Matrix Approach (\ETMA) for visible-light interface scattering is outlined.
}}

\vspace{\baselineskip}

\noindent\textcolor{white!90!black}{%
\fbox{\parbox{0.975\linewidth}{%
\textcolor{white!40!black}{\begin{tabular}{lr}%
  \begin{minipage}{0.6\textwidth}%
    {\small Copyright attribution to authors. \newline
    This work is a submission to SciPost Physics. \newline
    License information to appear upon publication. \newline
    Publication information to appear upon publication.}
  \end{minipage} & \begin{minipage}{0.4\textwidth}
    {\small Received Date \newline Accepted Date \newline Published Date}%
  \end{minipage}
\end{tabular}}
}}
}


\vspace{10pt}
\noindent\rule{\textwidth}{1pt}
\tableofcontents
\noindent\rule{\textwidth}{1pt}
\vspace{10pt}



\section{Introduction}
\label{sec:intro}


Scattering experiments are a versatile tool for the investigation of soft matter interfaces~\cite{COCIS,CORRELLI}. A variety of radiation types spanning X-rays, neutrons, and visible light provide complementary contrast mechanism which access different aspects of an interface layer~\cite{Renaud2009,Daillant2009,SINHA1988,Russell1990,EWDLS1986}. The different wavelengths display interface properties at different length scales. As scattering investigations are indirect measurements, a reliable theoretical description of the scattering processes in an interface profile is indispensable. Ideally, one would expect here a best practice approach among scientists which employ different types of radiation. Instead, the different traditions of diverse scientific communities have produced different theoretical frameworks with incompatible nomenclatures, making interdisciplinary exchange difficult. For an improvement, a strict separation of indiscriminate physics expectations and accurate mathematical content of a theory is useful. A naive physics based viewpoint might claim a direct transfer of an X-ray theory to visible light. The background of such an expectation is founded in the electromagnetic spectrum, which contains X-rays as well as visible light. This viewpoint pushes the burden of proof of what is different in a light scattering theory compared to the X-ray approach outside the X-ray community's responsibility. However, scattering in an interface layer inevitably needs the treatment of electromagnetic radiation in matter. It is this involvement where the mechanisms and their handed down theoretical conceptions are different for X-rays and light. From a mathematical viewpoint, different approaches need to be considered separately up to the point where a mathematical proof shows equality. So, the simplistic claim that an X-ray theory is suitable for visible light as well is not justified without such a proof. A blocking of a new interface light scattering theory by a false priority claim under the assumption that it necessarily has to be equivalent to X-ray theories is mathematically not admissible. Our analysis leads to a suggestion which connects the regimes for X-rays and visible light properly. It is different from prevailing X-ray theories.


The central theoretical tool of the X-ray interface scattering community is the Distorted Wave Born Approximation (\DWBA)~\cite{Renaud2009,Daillant2009,SINHA1988,Dietrich1995}. The \DWBA\ is presented in standard references with ``X-ray'' in the title, and discussions are confined to the X-ray regime. However, the approximations and scope of the theory are not stated explicitly. This creates an ambiguity: the \DWBA\ literature can be read either as implicitly restricting the theory to X-rays by virtue of context, or as claiming general validity for all electromagnetic radiation. The absence of a clear statement either way -- which we term the \emph{formidable ambiguity criterion} (\FAC) -- is a genuine problem when one attempts to develop theoretical tools for visible-light interface scattering, since it makes mathematical comparison with the \DWBA\ impossible in principle.

\begin{figure}[htbp]
\centering
\includegraphics[width=0.55\textwidth]{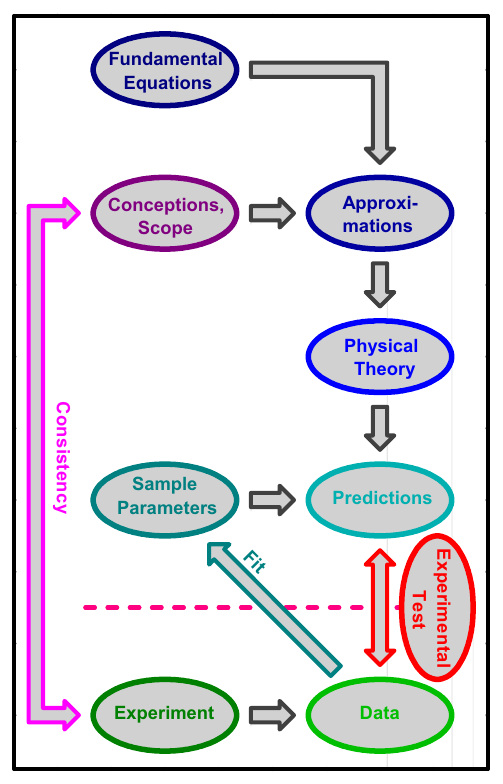}
\caption{Framework for critical analysis of a physical theory. Starting from fundamental equations, scope and approximations define the physical theory, which makes predictions compared with experimental data. Consistency between the experiment and the scope of the theory (pink arrows) is a necessary condition for physical understanding.}
\label{fig:scope}
\end{figure}

The present paper addresses this problem by providing an explicit analysis of the \DWBA's scope and approximations. Figure~\ref{fig:scope} illustrates the general framework for a critical analysis of any physical theory: starting from fundamental equations, a theory defines its scope (conceptions and approximations) and makes predictions that must be compared with experimental data from within that scope. 
Consistency between scope and experiment is essential; a theory that happens to describe data outside its scope does so by coincidence, not by physical understanding. Fitting parameters obtained from a theory applied outside its scope are not reliably comparable with independent measurements. The diagram furthermore clarifies the complementary nature of theory and experiment, which access the crucial experimental test from opposite sides.

In Figure~\ref{fig:scope}, the insufficiently discussed conceptions and approximations of the \DWBA\ constitutes the \FAC. Theoretical work is based on mathematical manipulations. Except for highly idealized cases, a full mathematical solution for arbitrary conditions cannot be achieved. It is thus unavoidable to involve approximations which fit to the target of the theoretical description. The resulting theory is restricted to experiments which meet the approximations. While the resulting domain of a theory is usually clearly indicated, the \FAC\ within the \DWBA\ literature obscures its scope. Beside this issue on the theory side, the \FAC\ affects also the experimental side, as the necessary consistency is not clear. 

Unfortunately, the not clarified \DWBA\ scope reproducibly leads to scientific disputes. From the perspective of X-ray researchers, the \DWBA\ is a well accepted theory which is cast into textbooks. The \FAC\ does not matter as long as one remains in the X-ray regime. It is comfortably overlooked. Reproducibly, any presentation of an interface scattering theory for visible light raises resolute resistance from some X-ray researchers, who consider the issue fully solved within their \DWBA. Such a point of view blithely ignores the general physics setting, that the question if a theory is useful can be exclusively clarified by experimental tests. A naive transfer of satisfactory \DWBA\ predictions for X-rays to visible light experiments is not conform with critical science. For reference in future work, we address such a push-back of new theoretical approaches for visible light scattering at interfaces as Mezger effect. The usual clarification of disputes by evidence from the literature cannot be achieved because of the \DWBA's \FAC. A mathematical comparison of different interface scattering theories is not possible, as long as the conception and the approximations of the \DWBA\ are not properly stated. So, there is an awful deadlock for a genuine interface light scattering theory by the rather poor \DWBA\ literature situation, in particular by the \FAC. The present work provides a clarification and differentiation.

For the \DWBA, the analysis shows that the fundamental equations differ from those of optical scattering theory. While optics is built on the macroscopic Maxwell equations (\MAME) with constitutive relations, the X-ray literature employs a microscopic picture based on electron densities and microscopic currents. This distinction -- not the wavelength itself -- is the root cause of the incompatibility between the two frameworks. There is an unconcerned intermixing of the microscopic Maxwell equations (\MIME) with the \MAME\ in the \DWBA. The muddle does not matter for X-rays, as the refractive index deviation from the vacuum value $1$ is minuscule in this wavelength regime. The setting is addressed as Small Refractive Index Deviation (\SRID). The central role of the \SRID\ as essential approximation of the \DWBA\ which enables the context swapping between \MIME\ and \MAME\ is not discussed in the \DWBA\ literature. While there is no effect for X-rays within experimental accuracy, a transfer of the \DWBA\ intermixing approach to the optical regime with substantial refractive index deviations from $1$ makes no sense.



The paper is organized as follows. Section~\ref{sec:SRID} discusses the \SRID\ condition and the fundamentally different electromagnetic foundations of X-ray and optical scattering theories. Section~\ref{sec:math} analyses two specific mathematical inconsistencies in the standard \DWBA\ derivation. Section~\ref{sec:ETMA} outlines the extended transfer matrix theory \ETMA\ as a rigorous alternative for visible-light interface scattering. Section~\ref{sec:discussion} discusses the scope of both theories and experimental tests. Section~\ref{sec:conclusions} states conclusions.

\section{Electromagnetic Foundations and the \SRID\ Condition}
\label{sec:SRID}

The incompatibility between X-ray and optical interface scattering theories does not arise from the wavelength difference alone. It arises from the use of fundamentally different electromagnetic descriptions.

\subsection{Visible Light in Matter}
\label{subsec:VisibleLight}

In optics, the relevant wavelengths are large compared with atomic spacings, and photon energies are well below ionization thresholds. The electric field of a light wave induces a polarization of the medium: bound electrons in atoms and molecules are displaced, creating oscillating dipole moments. A light wave in a medium is a coupled excitation of the field and the polarization of the medium. This physics is captured by the macroscopic Maxwell equations (\MAME):
\begin{equation}
\begin{array}{lclclcl}
\nabla \cdot \mathbf{D} & \!=\! & \RhoFree, & \quad &
\nabla \cdot \mathbf{B} & \!=\! & 0, \\[2mm]
\nabla \times \mathbf{E} & \!=\! & -\frac{\partial \mathbf{B}}{\partial t}, & \quad &
\nabla \times \mathbf{H} & \!=\! & \JFree + \frac{\partial \mathbf{D}}{\partial t}.
\end{array}
\label{Eq:MAME}
\end{equation}
Here, $\mathbf{E}$, $\mathbf{B}$, $\mathbf{D}$, $\mathbf{H}$, $\RhoFree$, and $\JFree$, are the electric field, the magnetic induction, the electric displacement, the magnetic field, the  density of free charges and the density of free currents, respectively. A closure of the system of equations (\ref{Eq:MAME}) requires the constitutive relations (\CRs)
\begin{equation}
\mathbf{D} = \EpsNull\EpsRel\mathbf{E}, \quad\quad
\mathbf{B} = \MuNull\Muriel\mathbf{H}, \quad\quad
\JFree = \sigma\mathbf{E}.
\label{Eq:ConstRel}
\end{equation}
Here, $\EpsNull$, $\MuNull$, $\sigma$, $\EpsRel$, and $\Muriel$, are the permittivity of vacuum, the permeability of vacuum, the conductivity, the relative permittivity and the relative permeability, respectively. The latter is essentially $1$ for optics and X-rays, so the factor is often omitted. As microscopic counterpart, the \MIME\ are written as 
\begin{equation}
\begin{array}{lclclcl}
\nabla \cdot \mathbf{e} & \!=\! & \frac{\rho}{\EpsNull}, & \quad &
\nabla \cdot \mathbf{b} & \!=\! & 0, \\[2mm]
\nabla \times \mathbf{e} & \!=\! & -\frac{\partial \mathbf{b}}{\partial t}, & \quad &
\nabla \times \mathbf{b} & \!=\! & \MuNull \left(\mathbf{j} + \EpsNull\frac{\partial \mathbf{e}}{\partial t}\right).
\end{array}
\label{Eq:MIME}
\end{equation}
Here, $\mathbf{e}$, $\mathbf{b}$, $\rho$, and $\mathbf{j}$ are the microscopic electric field, the microscopic magnetic induction, the microscopic charge density and the microscopic current density. Despite the similarity in the form, the systems (\ref{Eq:MAME}) and (\ref{Eq:MIME}) involve different quantities and thus describe different physical aspects. Within the diagram of Figure \ref{fig:scope}, there is a essential difference of a setting which uses a macroscopic approach and inserts the combination of (\ref{Eq:MAME}) and (\ref{Eq:ConstRel}) as fundamental equations on one hand and a microscopic approach with (\ref{Eq:MIME}) as basis on the other hand.

The optics literature employs a hierarchical approach, where the \MIME\ (\ref{Eq:MIME}) are used for a derivation of the \MAME\ (\ref{Eq:MAME}). Starting point are the assignments $\langle\mathbf{e}\rangle=\mathbf{E}$ and $\langle\mathbf{b}\rangle=\mathbf{B}$: the spatial averages of the microscopic fields which characterize the forces on charges are identified with macroscopic fields. Such an approach is called mean field theory within statistical mechanics. The latter discipline provides a good understanding of the scope of a mean field theory. It works well except for situations with large density fluctuations, e.g.\ close to a second order phase transition. It is interesting to realize that at these conditions the basis of the \MAME\ might break down too. On the other hand, the mean field approach works well for moderate density variations. 

A first approach for the derivation of the \MAME\ from the \MIME\ \cite{Jackson} dates back to 1906 and goes back to Lorenz. It makes a distinction between charges bound in molecules and free charges. The electrical field generates a polarization $\mathbf{P}$ in the molecules according to $\mathbf{P}=\EpsNull(\EpsRel-1)\mathbf{E}$. It covers the behavior of all bound charges, which do not occur in the \MAME\ any more explicitly. The free charges $\RhoFree$ are ions in solution or mobile electrons in a metal. Their motion establishes $\JFree$. The symbols and nomenclature in (\ref{Eq:MAME}) originates from this background. An issue of this first approach is related to the \CRs. The derivation uses the simple multiplication form (\ref{Eq:ConstRel}) in the time domain. However, the simple form is appropriate to the frequency domain. It is a well established experimental finding, that the refractive index (\RI) which is directly related to $\EpsRel$ depends on the light frequency. A proper transfer of (\ref{Eq:ConstRel}) from the frequency domain to the time domain requires the application of the Faltungstheorem, which yields a convolution in the time domain, different from (\ref{Eq:ConstRel}).

A more satisfactory approach derives the \MAME\ directly in the frequency domain, avoiding the convolution issue altogether. Taking the temporal Fourier transform of the \MIME\ (Eq.~\ref{Eq:MIME}), one replaces $\partial/\partial t \to -i\omega$ and works with the complex amplitudes $\tilde{\mathbf{e}}(\mathbf{r},\omega)$, $\tilde{\mathbf{b}}(\mathbf{r},\omega)$, etc. A spatial ensemble average $\langle\cdot\rangle$ then defines the macroscopic fields: $\mathbf{E}(\mathbf{r},\omega)=\langle\tilde{\mathbf{e}}(\mathbf{r},\omega)\rangle$ and $\mathbf{B}(\mathbf{r},\omega)=\langle\tilde{\mathbf{b}}(\mathbf{r},\omega)\rangle$~\cite{MazurNijboer1953,VanKranendonkSipe1977}. This statistical averaging, analogous to a mean-field approach in statistical mechanics, was introduced by Mazur and Nijboer~\cite{MazurNijboer1953} and developed further by van Kranendonk and Sipe~\cite{VanKranendonkSipe1977}. The averaged \MIME\ take the form of the \MAME\ (Eq.~\ref{Eq:MAME}), with the macroscopic polarization $\mathbf{P}(\mathbf{r},\omega)$ and magnetization $\mathbf{M}(\mathbf{r},\omega)$ arising from the averaging of the microscopic multipole densities. 

In the frequency domain, the \CRs\ take the simple multiplicative form~\cite{VanKranendonkSipe1977,IlinskiiKeldysh1994}:
\begin{equation}
    \tilde{\mathbf{D}}(\mathbf{r},\omega) = \EpsNull\EpsRel(\omega)\,\tilde{\mathbf{E}}(\mathbf{r},\omega),
    \label{Eq:CRFreq}
\end{equation}
where $\EpsRel(\omega)$ is the frequency-dependent relative permittivity. This is the natural domain for the \CRs: causality requires that the response function in the time domain be retarded (i.e.\ $\EpsRel(t-t')=0$ for $t'> t$), so that equation~(\ref{Eq:CRFreq}) in the frequency domain corresponds to a convolution $\mathbf{D}(t)=\EpsNull\int_{-\infty}^{t}\EpsRel(t-t')\mathbf{E}(t')\,\mathrm{d}t'$ in the time domain via the convolution theorem~\cite{LandauLifshitzElectrodyn,Nussenzveig1972}. The analyticity of $\EpsRel(\omega)$ in the upper complex half-plane, enforced by causality, gives rise to the Kramers--Kronig dispersion relations connecting the real and imaginary parts of $\EpsRel(\omega)$~\cite{LandauLifshitzElectrodyn}. The frequency-domain approach is the basis of modern treatments of dispersive media~\cite{IlinskiiKeldysh1994,MelroseMcPhedran1991} and is the appropriate framework for any scattering theory that requires frequency-dependent material response.

The standard model used in optics assumes a non-absorbing dielectric medium with $\RhoFree=0$, $\JFree=0$, and $\sigma=0$. Absorbing media can be treated for non-zero angular frequency $\omega$ by the introduction of a generalized dielectric displacement $\mathbf{D}^{(g)}=\mathbf{D}+i\JFree/(-i\omega)$ (we employ the $\exp(-i\omega t)$ convention for the dependence on the time $t$) \cite{BedeauxVlieger}. The elimination of $\mathbf{D}$ in favor of $\mathbf{D}^{(g)}$ removes $\RhoFree$ and $\JFree$ from the \MAME\ (\ref{Eq:MAME}) and combines two \CRs\ (\ref{Eq:ConstRel}) as $\mathbf{D}^{(g)}=\EpsNull\EpsRel^{(g)}\mathbf{E}$ with the generalized relative permittivity
$\EpsRel^{(g)}=\EpsRel+i\sigma\EpsNull^{-1}\omega^{-1}$. The mathematical transformation without further physics assumptions maps the \MAME\ with charges and currents and a real valued relative permittivity $\EpsRel$ to charge-free and current-free \MAME\ with complex valued relative permittivity $\EpsRel^{(g)}$. While $\mathbf{P}$ and $\mbox{Re}(\mathbf{D}^{(g)})$ contain charge oscillations in-phase with $\mathbf{E}$, the free currents hidden in $\mbox{Im}(\mathbf{D}^{(g)})$ involve charge oscillations out-of-phase relative to $\mathbf{E}$. The later cause light absorption. The physical interpretation of $\RhoFree$ and $\JFree$ in this framework is different from above. It can be concluded, there is an inconsistency in the literature. As, however, optics calculations assume $\RhoFree=0$ and $\JFree=0$ for dielectric material, the different interpretations do not matter.\\

\subsection{X-rays in Matter}
\label{subsec:Xray}

For X-rays, the wavelength is comparable with atomic spacings and higher photon energies can free electrons from the molecular assembly. The theory starts from the electron density in matter and assumes the behavior of free electrons. Within the diagram of Figure \ref{fig:scope}, this setting is a conception and an approximation. Even though the electrons can be freed, they do suffer electromagnetic interactions with the other charges. The success of the assumption might rely on the smearing of electron locations and the averaging of interactions in different directions with a multitude of other charges. The X-ray literature easily suspends this assumption when it is convenient, e.g.\ for the theoretical description of Anomalous Small Angle X-ray Scattering (\ASAXS)~\cite{ASAXS_Stuhrmann1985,ASAXS_Svergun1994}. This technique varies the X-ray energy in a region around the binding energy of an inner electron. In general, a clearer indication and discussion of physical assumptions in the X-ray literature would be desirable.

When total external reflection is to be described, the X-ray literature introduces a \RI\ via Thomson scattering~\cite{deBergevin2009}. The field of the illuminating radiation shakes the assumed free electrons. These oscillating charges emit secondary radiation. The interference of the latter in forward scattering direction with the illuminating radiation modifies the propagation in matter. For a theoretical description of this modification, a \RI\ $n$ is introduced. 
The derivation route and the underlying physical picture differ from the optical case. 

At this point we suggest a concept which provides a proper connection to optical theories. It transforms the hierarchical approach of visible light electromagnetism in matter to X-rays. The Thomson scattering mechanism is considered as the microscopic picture to derive the \MAME\ with the usual setting $\EpsRel=n^2$. A proper scattering theory is then based exclusively on the \MAME\ with no currents and no charges in combination with the \CRs, like in optics. The idea is that there is a general modification of the speed of light in media by a \RI\ at all wavelengths. Scattering is caused by fluctuations of $\EpsRel$.

The \DWBA, as formulated in standard references~\cite{Renaud2009,Daillant2009,SINHA1988,Dietrich1995} follows a different path. The discussion is heavily based on oscillating microscopic charges and currents, in contrast to optical theories with $\RhoFree=0$ and $\JFree=0$. The \DWBA\ does not maintain a strict separation between \MIME\ and \MAME. The reciprocity theorem (see Section~\ref{sec:math}) is derived from the \MAME\, where subsequently $\JFree$ is identified with microscopic currents. Such an approach makes no sense for an optical theory. The \DWBA\ treats the polarization via a fictitious point dipole in vacuum, and mixes microscopic and macroscopic descriptions throughout. This approach is working within the X-ray regime, where the consequences of the mixing are negligible. It is not compatible with the optical framework built on the \MAME. As it has been a particular matter of a debate, we explicitly remark that the foundation of the X-ray \RI\ based on the free electrons assumption excludes birefringence, which is caused by anisotropic polarizability. So, the \DWBA\ cannot describe the scattering in a birefringent interface layer.

\subsection{The \SRID\ Condition}
\label{subsec:SRID}

The key quantitative distinction between the two regimes is the small refractive index deviation (\SRID) condition. For X-rays, the \RI\ is conventionally written as $n=1-\delta+i\beta$, where $\delta\sim10^{-6}$ and $\beta$ is typically an order of magnitude smaller~\cite{deBergevin2009}. The deviation from unity is extremely small. At an interface between two materials, the contrast $\Delta n = n_1 - n_2$ is of the same order.

For visible light, \RI\ values lie typically in the range $1.3$ to $1.9$. The contrast at a water--air interface ($n_\text{water}\approx1.333$, $n_\text{air}=1$) is of order $0.3$ -- approximately $10^6$ times larger than typical X-ray contrasts. This is not a quantitative difference of degree; it is a qualitative distinction that invalidates several approximations underlying the \DWBA.

The \SRID\ condition has three important consequences for X-ray interface scattering, all of which break down for visible light:

\begin{enumerate}
\item 
\textbf{RI gradients.} Refractive index gradients $\mathrm{d}n/\mathrm{d}z$ at an interface scale with $|n-1|$. For X-rays, these gradients are negligibly small compared with typical scattering vector magnitudes $q \sim 2\pi n/\lambda$. The \DWBA\ does not include gradient scattering contributions, which is unproblematic for X-rays but constitutes a significant omission for visible-light scattering.
\item 
\textbf{Mixing of micro- and macroscopic descriptions.} For X-rays, $n \approx 1$ implies that the vacuum speed of light $c$ and the in-medium speed $c/n$ are nearly identical. The consequences of mixing \MIME\ and \MAME\ are therefore numerically negligible. For visible light, where $n$ differs substantially from $1$, such mixing introduces uncontrolled errors.
\item \textbf{Polarization.} The \DWBA\ treats polarization within the small-angle approximation appropriate for Grazing Incidence Small Angle X-ray Scattering (GISAXS)~\cite{Renaud2009}. This approximation predicts equal scattering for both polarization states. For visible-light scattering with large RI contrast and significant RI gradients, polarization-dependent effects are expected to be substantial. A proper theoretical description is required for advanced experiments like scattering ellipsometry~\cite{Germer2000,CORRELLI}.
\end{enumerate}

None of these limitations are stated explicitly in the standard \DWBA\ literature~\cite{Renaud2009,Daillant2009,SINHA1988,Dietrich1995}. The theory is described as being ``for X-rays'' in its titles and context, but without a clear analysis of which approximations make it specific to X-rays. This constitutes what we term the \emph{formidable ambiguity criterion} (\FAC): the \DWBA\ literature is ambiguous as to whether the theory is restricted to X-rays by context, or claims general applicability. The \FAC\ prevents any rigorous mathematical assessment of whether a new visible-light scattering theory is independent of, consistent with, or in contradiction to the \DWBA.

\section{Mathematical Analysis of the \DWBA}
\label{sec:math}

\subsection{The Reciprocity Theorem}
\label{subsec:rt}

A central element of the \DWBA\ is the reciprocity theorem (RT), which allows the propagation of radiation from a scattering event to the detector to be replaced by the inverse propagation from detector to scattering event. The conventional derivation~\cite{Daillant2009} proceeds from a first reciprocity relation following Landau and Lifshitz (\S69 of~\cite{LandauLifshitzElectrodyn}):
\begin{equation}
    \int \mathrm{d}^3r \, \mathbf{E}_B \cdot \JFree_A =
    \int \mathrm{d}^3r \, \mathbf{E}_A \cdot \JFree_B,
    \label{eq:rt1}
\end{equation}
where $\mathbf{E}$ and $\JFree$ denote electric field and free current density for two systems $A$ and $B$.

Equation~(\ref{eq:rt1}) involves free current densities, which are zero for dielectric materials. For the dielectrics of primary interest in visible-light optics, equation~(\ref{eq:rt1}) is satisfied trivially and provides no useful information for a scattering theory. Even when free currents are present, they do not generally coincide with the sources of optical scattering contrast, which arise from the dielectric properties of the scattering objects. A scattering theory for visible light founded on free-current reciprocity contains a conceptual mismatch at its starting point.

The \DWBA\ derivation continues by introducing the substitution $\mathbf{P} = \JFree/(i\omega)$, where $\mathbf{P}$ is described as a ``dipole density'', and inserting into equation~(\ref{eq:rt1}) to obtain:
\begin{equation}
    \int \mathrm{d}^3r \, \mathbf{E}_B \cdot \mathbf{P}_A =
    \int \mathrm{d}^3r \, \mathbf{E}_A \cdot \mathbf{P}_B.
    \label{eq:rt2}
\end{equation}
This substitution does not appear as a numbered equation in the original literature~\cite{Daillant2009}, and the term ``dipole density'' deviates from the standard meaning of $\mathbf{P}$ as macroscopic polarization in optics. No statement is made that the dipole density equals the polarization, nor are differences discussed.

Equation~(\ref{eq:rt2}) is in fact correct, but for a reason unrelated to the derivation above. A direct one-line proof requires only the linearity of the constitutive relation 
$\mathbf{D}=\EpsNull\EpsRel\mathbf{E}=\EpsNull\mathbf{E}+\mathbf{P}$:
\begin{equation}
    \mathbf{E}_B \cdot \mathbf{P}_A
    = \mathbf{E}_B \cdot (\EpsNull\EpsRel\mathbf{E}_A
    - \EpsNull \mathbf{E}_A)
    = \mathbf{E}_A \cdot (\EpsNull\EpsRel\mathbf{E}_B
    - \EpsNull\mathbf{E}_B)
    = \mathbf{E}_A \cdot \mathbf{P}_B.
    \label{eq:rt_proof}
\end{equation}
Integration of both sides over the sample volume immediately yields equation~(\ref{eq:rt2}). This derivation requires no assumptions about free currents, no microscopic picture of oscillating charges, and is fully compatible with the \MAME\ framework. The fact that the conventional \DWBA\ derivation arrives at the correct result via an unjustified route is a symptom of the \MIME/\MAME\ mixing identified in Section~\ref{sec:\SRID}: within the X-ray regime where $n \approx 1$, the numerical consequences are negligible and the error is invisible.

\subsection{The Point-Dipole Assumption at the Detector}
\label{subsec:dipole}

To obtain a practical scattering amplitude, the \DWBA\ inserts into equation~(\ref{eq:rt2}) the polarization of a fictitious unit dipole in the direction of the unit vector $\hat{e}$ at the detector location $\mathbf{R}$:
\begin{equation}
    \mathbf{P}_A(\mathbf{r}) = \delta(\mathbf{R} - \mathbf{r})\hat{e}.
    \label{eq:dipole}
\end{equation}
Here, $\delta(.)$ designates the delta-function.
This assumption has two formal problems. First, the detector is located in vacuum, where $\EpsRel\equiv1$ by definition. The macroscopic polarization at any point in vacuum is therefore $\mathbf{P}(\mathbf{R})=\EpsNull(\EpsRel-1)\mathbf{E}(\mathbf{R})=0$. A non-zero point-like polarization in vacuum cannot be justified within the \MAME\ and \CRs. Equation~(\ref{eq:dipole}) is an ad-hoc postulate. It cannot be integrated in the diagram of Figure \ref{fig:scope}, as it is an extra assumption which is mathematically not compatible to the fundamental equations.

Second, equation~(\ref{eq:dipole}) assigns units of inverse volume (m$^{-3}$) on the right side to $\mathbf{P}_A$ on the left side, while the macroscopic polarization $\mathbf{P}_B$ in the scattering volume has units of C\,m$^{-2}$. The two sides of equation~(\ref{eq:rt2}), after insertion of equation~(\ref{eq:dipole}), involve quantities of different dimensions. This dimensional inconsistency is not noted or discussed in the \DWBA\ literature.

The motivation for equation~(\ref{eq:dipole}) derives from the X-ray picture of matter as a collection of free electrons emitting dipole radiation -- a microscopic description. Within the macroscopic framework of the \MAME, the introduction of a fictitious dipole in vacuum is not admissible. In a correctly formulated scattering theory, the detector simply measures the intensity of the electromagnetic field. In particular, it does not host an extra polarization.

\subsection{Summary of Inconsistencies}
\label{subsec:summary_math}

The two inconsistencies identified above -- the detour via free-current reciprocity and the fictitious vacuum dipole -- share a common origin: the uncontrolled mixing of \MIME\ and \MAME\ descriptions. For X-rays, where $n \approx 1$ and the distinction between microscopic and macroscopic descriptions is numerically irrelevant, the inconsistencies do not affect the practical utility of the \DWBA. For visible light, where $n$ differs substantially from unity, they introduce errors that cannot be controlled.

\section{The Extended Transfer Matrix Approach}
\label{sec:ETMA}

An alternative theoretical framework for visible-light interface scattering, the extended transfer matrix approach (\ETMA), has been developed and presented at a conference~\cite{TalkETMA2019}. The \ETMA\ is built exclusively on the \MAME\ and \CRs. It uses $8\times8$ transfer matrices to conjointly describe light propagation and scattering in layered \RI\ profiles. An essential pillar is the layer scalar product~\cite{TRAMATSINC} for the calculation of scattering within a layer. Interface-bound fluctuations (\IBF)~\cite{SOFTINTERFACES} constitute the scattering in an interface \RI\ profile. Fluctuations affect the local \RI-gradient. For p-polarization, these gradient effects modify the scattering amplitude. Their treatment in the \ETMA\ provides a reliable description of polarization-dependent scattering at large \RI\ contrast and for birefringent media.

The \ETMA\ does not employ the Green's function formalism, does not introduce fictitious dipoles in vacuum, and does not rely on the \SRID\ condition. When \SRID\ conditions are inserted into the \ETMA, it should be possible to recover a version of the \DWBA\ with properly stated approximations, replacing the current inconsistent derivation while leaving unchanged the practical application for X-ray experiments.

Experimental evidence for the physical reality of amplitude divergence at critical conditions (\CC) 
has been obtained from \EWDLS\ measurements on nematic wetting layers~\cite{Sigel1997,Sigel2000} and capillary waves at a water surface~\cite{Stocco2009}. The companion paper outlines how these data can be described by the \ETMA~\cite{SigelDivAmp}. The discussion shows, that the \CC\ mechanism provides the dominant scattering contribution for the scattering of evanescent waves created from visible coherent laser light, while it is basically absent for the case of incoherent synchrotron radiation and \SRID\ conditions of X-rays. As this dominant mechanism is not covered by the \DWBA, this theory is useless for visible light.

Unfortunately, a publication of the \ETMA\ is rather difficult so far, as the \FAC\ makes a discussion of differences to the \DWBA\ impossible. A proper definition of terms is the very fundament of any scientific activity which necessarily has to precede theoretical work in physics. The vagueness of the \FAC\ decomposes the scientific language which is required for physics discussions. The clarification of terms and conceptions within this work paves the way for the \ETMA.

\section{Discussion}
\label{sec:discussion}

\subsection{Scope and Experimental Tests}
\label{subsec:scope}

Based on the preceding analysis, the scopes of the \DWBA\ and \ETMA\ can be characterized precisely.

The \DWBA\ is a productive and well-validated theory for X-ray interface scattering under \SRID\ conditions. Within this scope -- \GISAXS\ and related X-ray experiments -- the mathematical inconsistencies identified in Section~\ref{sec:math} as well as the absence of the \CC\ mechanism (see section \ref{sec:ETMA}) do not prevent useful predictions, because the \SRID\ condition and the incoherent illumination makes their numerical consequences negligible. The \DWBA\ is not adequate as a theoretical framework for visible-light interface scattering in a total reflection geometry.

The \ETMA\ is the appropriate framework for interface scattering based on the \MAME\ and \CRs. This combination includes visible light and -- within the hierarchical approach suggested in section \ref{subsec:Xray} -- also X-rays. Its predictions differ from the \DWBA\ in three experimentally accessible ways: (i) in a detailed comparison of amplitude ratio and relative phase shift of p- and s-polarization by scattering ellipsometry, the \ETMA\ considers gradient scattering contributions, which are not contained in the \DWBA; (ii) the \ETMA\ predicts amplitude divergence at critical conditions that produces a dominant scattering contribution absent from the \DWBA; (iii) while the \ETMA\ makes predictions for birefringent samples, it is unclear and not covered by the literature yet, how the \DWBA's free microscopic electrons procedures can be adapted to birefringent samples.


\subsection{On the Importance of Explicit Scope Statements}
\label{subsec:importance}

The framework illustrated in Figure~\ref{fig:scope} makes clear why an explicit statement of scope and approximations is scientifically important. A theory applied outside its scope may still fit experimental data, because free fitting parameters can absorb discrepancies. Furthermore, interface scattering is typically not performed with absolute intensity measurements, so there is a free scaling parameter in the fitting procedure which can mask a quantitative disagreement of a theory. An agreement achieved in such a way does not constitute physical understanding. Parameters obtained by fitting a theory outside its scope are not reliably comparable with independent measurements of the same physical quantities.

For the \DWBA, the absence of an explicit scope statement has practical consequences. It has made it impossible, in prior scientific discussions, to resolve the question of whether the \DWBA\ applies to visible-light interface scattering by reference to the literature. The present analysis -- by identifying the \SRID\ condition as the operative approximation and demonstrating specific mathematical inconsistencies that arise when it is violated -- provides the basis for such a literature reference.

The practical consequence for the X-ray community is minimal: the \DWBA\ remains valid and productive for \GISAXS\ and related experiments. The consequence for the visible-light interface scattering community is that the field is now open to proper theoretical development, with a clear analysis establishing that the \DWBA\ does not constitute an adequate pre-existing theory for this regime.

\section{Conclusions}
\label{sec:conclusions}

The main conclusions of this paper are as follows:

\begin{enumerate}
\item The \DWBA\ and optical scattering theories rest on fundamentally different electromagnetic foundations. The \DWBA\ is built on a microscopic description using electron densities and microscopic currents, while optical scattering theory must be built on the macroscopic Maxwell equations and constitutive relations. This distinction, not the wavelength per se, is the root cause of the incompatibility between the two frameworks.
\item The small refractive index deviation (\SRID) condition -- $|n-1| \sim 10^{-6}$ for X-rays versus $|n-1| n \sim 0.5$ for visible light, a difference of six orders of magnitude -- is the essential implicit prerequisite of the \DWBA. It makes \RI\ gradients negligible, renders the \MIME/\MAME\ mixing numerically irrelevant, and allows a uniform treatment of both polarization states.
\item The \DWBA\ literature does not state the \SRID\ condition explicitly as a prerequisite, creating a formidable ambiguity criterion (\FAC) about the theory's domain of applicability. This prevents rigorous mathematical comparison with alternative theories.
\item The conventional derivation of the reciprocity theorem in the \DWBA\ involves an unjustified detour via free current densities, which are zero for optical dielectrics. A direct one-line proof using only the constitutive relation is provided.
\item The introduction of a fictitious unit dipole at the detector in the \DWBA\ is incompatible with the constitutive relations of classical electrodynamics and leads to a dimensional inconsistency.
\item The extended transfer matrix approach (\ETMA) provides a rigorous framework for visible-light interface scattering built exclusively on the \MAME\ and \CRs. Scattering ellipsometry and measurements at critical conditions in a total reflection geometry provide experimental tests distinguishing the two theories.
\item Insertion of \SRID\ conditions into the \ETMA\ is expected to yield a properly derived version of the \DWBA\ with explicit approximations, clarifying the relationship between the two frameworks.
\end{enumerate}

\section*{Data Availability}

The scientific content, arguments, and calculations of this work were developed by the author. In preparing the manuscript, writing assistance was provided by the artificial intelligence assistant Claude Sonnet~4.6 (Anthropic). The AI's role was limited to organizing existing material, ensuring clarity of presentation, and achieving appropriate academic tone. All scientific responsibility rests with the author. For documentation purposes, previous versions of this manuscript which are used as input for the \AI\ are available as supplementary material \cite{SupplementaryMaterial}.

\section*{Acknowledgements}

The author thanks Matthias Fuchs for continued interest in this work and for the opportunity to attend the group seminar. Helpful discussions with Gert Strobl are gratefully acknowledged. The inspiration and practical help for the use of artificial intelligence provided by Cornelia Wetzel is highly appreciated. This manuscript was drafted with writing assistance from the AI assistant Claude Sonnet~4.6 (Anthropic); all scientific content was developed by the author. We thank Helgard Sigel for her kind hospitality.

\paragraph{Funding information}
This work has been done without external funding.

\bibliography{DWBAAnalysisSciPost}

@article{Germer2000,
  author  = {Germer, Thomas A.},
  title   = {Measurement of Roughness of Two Interfaces of a Dielectric Film by Scattering Ellipsometry},
  journal = {Phys.\ Rev.\ Lett.},
  volume  = {85},
  number  = {2},
  pages   = {349--352},
  year    = {2000},
  doi     = {10.1103/PhysRevLett.85.349}
}

@article{Russell1990,
  author  = {Russell, T. P.},
  title   = {X-ray and neutron reflectivity for the investigation of polymers},
  journal = {Materials Science Reports},
  volume  = {5},
  number  = {4},
  pages   = {171--271},
  year    = {1990},
  doi     = {10.1016/0920-2307(90)90009-D}
}

@article{MazurNijboer1953,
  author  = {Mazur, P. and Nijboer, B. R. A.},
  title   = {On the statistical mechanics of matter in an electromagnetic field. {I}: Derivation of the {M}axwell equations from electron theory},
  journal = {Physica},
  volume  = {19},
  pages   = {971--986},
  year    = {1953},
  doi     = {10.1016/S0031-8914(53)80109-6}
}

@incollection{VanKranendonkSipe1977,
  author    = {Van Kranendonk, J. and Sipe, J. E.},
  title     = {Foundations of the Macroscopic Electromagnetic Theory of Dielectric Media},
  booktitle = {Progress in Optics},
  editor    = {Wolf, E.},
  volume    = {15},
  pages     = {245--350},
  publisher = {North-Holland},
  address   = {Amsterdam},
  year      = {1977},
  doi       = {10.1016/S0079-6638(08)70480-3}
}

@book{IlinskiiKeldysh1994,
  author    = {Il'inskii, Yu. A. and Keldysh, L. V.},
  title     = {Electromagnetic Response of Material Media},
  publisher = {Plenum Press},
  address   = {New York},
  year      = {1994}
}

@book{MelroseMcPhedran1991,
  author    = {Melrose, D. B. and McPhedran, R. C.},
  title     = {Electromagnetic Processes in Dispersive Media: A Treatment Based on the Dielectric Tensor},
  publisher = {Cambridge University Press},
  address   = {Cambridge},
  year      = {1991}
}

@book{Nussenzveig1972,
  author    = {Nussenzveig, H. M.},
  title     = {Causality and Dispersion Relations},
  publisher = {Academic Press},
  address   = {New York},
  year      = {1972}
}

@article{ASAXS_Stuhrmann1985,
  author  = {Stuhrmann, H. B.},
  title   = {Resonance scattering in macromolecular structure research},
  journal = {Advances in Polymer Science},
  volume  = {67},
  pages   = {123--163},
  year    = {1985},
  doi     = {10.1007/BFb0017792}
}

@article{ASAXS_Svergun1994,
  author  = {Svergun, D. I. and Koch, M. H. J.},
  title   = {Anomalous small-angle {X}-ray scattering},
  journal = {Rep.\ Prog.\ Phys.},
  volume  = {66},
  pages   = {1735--1782},
  year    = {2003},
  doi     = {10.1088/0034-4885/66/10/R05}
}

@book{BedeauxVlieger,
author = {Bedeaux, Dick and Vlieger, Jan},
title = {Optical Properties of Surfaces},
publisher = {PUBLISHED BY IMPERIAL COLLEGE PRESS AND DISTRIBUTED BY WORLD SCIENTIFIC PUBLISHING CO.},
year = {2004},
doi = {10.1142/p327},
address = {},
edition   = {2nd},
URL = {https://www.worldscientific.com/doi/abs/10.1142/p327},
eprint = {https://www.worldscientific.com/doi/pdf/10.1142/p327}
}

@article{EWDLS1986,
  title = {Brownian dynamics close to a wall studied by photon correlation spectroscopy from an evanescent wave},
  author = {Lan, K. H. and Ostrowsky, N. and Sornette, D.},
  journal = {Phys.\ Rev.\ Lett.},
  volume = {57},
  issue = {1},
  pages = {17--20},
  numpages = {0},
  year = {1986},
  month = {Jul},
  publisher = {American Physical Society},
  doi = {10.1103/PhysRevLett.57.17},
  url = {http://link.aps.org/doi/10.1103/PhysRevLett.57.17}
}

@ARTICLE{COCIS,
  author = {Reinhard Sigel},
  title = {Light scattering near and from interfaces using evanescent wave and ellipsometric light scattering},
  journal = {Current Opinion in Colloid \& Interface Science},
  year = {2009},
  volume = {14},
  number = {6},
  pages = {426--437},
  doi = {10.1016/j.cocis.2009.08.004},
}

@article{SIGEL1997, 
author = {R. Sigel and G. Strobl}, 
journal = {Prog.\ Coll.\ Polym.\ Sci.}, 
pages = {187--190}, 
title = {Static and Dynamic Light Scattering from the Nematic Wetting Layer}, 
volume = {104}, 
year = {1997}
}

@article{SIGEL2000, 
author = {R. Sigel and G. Strobl}, 
journal = {J.\ Chem.\ Phys.}, 
number = {12}, 
pages = {1029--1039}, 
title = {Light scattering by fluctuations within a nematic wetting layer in a isotropic phase of a liquid crystal}, 
volume = {112}, 
year = {2000}
}

@article{Renaud2009,
title = "Probing surface and interface morphology with Grazing Incidence Small Angle X-Ray Scattering",
journal = "Surface Science Reports",
volume = "64",
number = "8",
pages = "255 - 380",
year = "2009",
issn = "0167-5729",
doi = "https://doi.org/10.1016/j.surfrep.2009.07.002",
url = "http://www.sciencedirect.com/science/article/pii/S0167572909000399",
author = "Gilles Renaud and R\´emi Lazzari and F\´ed\´eric Leroy"
}

@Inbook{Daillant2009,
author="Daillant, J. and Mora, S. and Sentenac, A.",
editor="Daillant, Jean and Gibaud, Alain",
title="Diffuse Scattering",
bookTitle="X-ray and Neutron Reflectivity: Principles and Applications",
year="2009",
publisher="Springer Berlin Heidelberg",
address="Berlin, Heidelberg",
pages="133--182",
isbn="978-3-540-88588-7",
doi="10.1007/978-3-540-88588-7_4",
url="https://doi.org/10.1007/978-3-540-88588-7_4"
}

@article{SINHA1988,
  title = {X-ray and neutron scattering from rough surfaces},
  author = {Sinha, S. K. and Sirota, E. B. and Garoff, S. and Stanley, H. B.},
  journal = {Phys. Rev. B},
  volume = {38},
  issue = {4},
  pages = {2297--2311},
  numpages = {0},
  year = {1988},
  month = {Aug},
  publisher = {American Physical Society},
  doi = {10.1103/PhysRevB.38.2297},
  url = {https://link.aps.org/doi/10.1103/PhysRevB.38.2297}
}

@article{DIETRICH1995,
title = "Scattering of X-rays and neutrons at interfaces",
journal = "Physics Reports",
volume = "260",
number = "1-2",
pages = "1 -- 138",
year = "1995",
issn = "0370-1573",
doi = "https://doi.org/10.1016/0370-1573(95)00006-3",
url = "http://www.sciencedirect.com/science/article/pii/0370157395000063",
author = "S. Dietrich and A. Haase"
}

@Inbook{deBergevin2009,
author="de Bergevin, F.",
editor="Daillant, Jean and Gibaud, Alain",
title="The Interaction of X-Rays (and Neutrons) with Matter",
bookTitle="X-ray and Neutron Reflectivity: Principles and Applications",
year="2009",
publisher="Springer Berlin Heidelberg",
address="Berlin, Heidelberg",
pages="1--57",
isbn="978-3-540-88588-7",
doi="10.1007/978-3-540-88588-7_1",
url="https://doi.org/10.1007/978-3-540-88588-7_1"
}

@Article{CORRELLI,
author ="Sigel, Reinhard",
title  ="Foundation of correlation ellipsometry",
journal  ="Soft Matter",
year  ="2017",
volume  ="13",
issue  ="6",
pages  ="1132-1141",
publisher  ="The Royal Society of Chemistry",
doi  ="10.1039/C6SM02285E",
url  ="http://dx.doi.org/10.1039/C6SM02285E"}

@book{Jackson,
  address = {New York, {NY}},
  author = {Jackson, John David},
  edition = {3rd ed.},
  isbn = {9780471309321},
  lccn = {538.3537.8},
  publisher = {Wiley},
  title = {Classical electrodynamics},
  url = {http://cdsweb.cern.ch/record/490457},
  year = 1999
}

@book{LandauLifshitzElectrodyn,
  address = {New York},
  author = {Landau, L. D. and Lifshitz, E. M.},
  publisher = {Pergamon},
  title = {Electrodynamics of Continuous Media},
  year = 1984
}

@MISC{SigelDivAmp,
  author = {Reinhard Eduard Sigel},
  title = {Divergence of light wave amplitudes in an interface layer at critical conditions},
  year         = 2026,
  howpublished = {{\it SciPost Physics}, submitted}  
}

@misc{TalkETMA2019,
  title        = {An Extended Transfer Matrix Approach to Calculate the Scattering of Light in an Interface Profile},
  author       = {Sigel, R.},
  year         = 2019,
  howpublished = {https://www.dpg-verhandlungen.de/year/2019/conference/regensburg/\\part/cpp/session/8/contribution/8?lang=en},
  note         = {Accessed on April 15th 2026}
}

@misc{SupplementaryMaterial,
  author       = {Sigel, R.E.},
  year         = 2026,
  doi          = {10.5281/zenodo.23169722},
  howpublished = {https://zenodo.org/records/23169722},
  note         = {Accessed on October 5th 2026}
}

@article{TRAMATSINC,
author = {Reinhard Sigel},
journal = {J. Opt. Soc. Am. A},
number = {12},
pages = {2142--2152},
publisher = {Optica Publishing Group},
title = {Light propagation in layered media in a total reflection geometry: a transfer matrix method using virtually linear basis functions to handle critical conditions},
volume = {39},
month = {Dec},
year = {2022},
url = {https://opg.optica.org/josaa/abstract.cfm?URI=josaa-39-12-2142},
doi = {10.1364/JOSAA.472361}
}

@Article{SOFTINTERFACES,
author ="Sigel, Reinhard",
title  ="Concepts for soft interfaces",
journal  ="Soft Matter",
year  ="2017",
volume  ="13",
issue  ="10",
pages  ="1940-1942",
publisher  ="The Royal Society of Chemistry",
doi  ="10.1039/C6SM02413K",
url  ="http://dx.doi.org/10.1039/C6SM02413K"}

@ARTICLE{STOCCO2009,
  author = {Antonio Stocco and Klaus Tauer and Stergios Pispas and Reinhard Sigel},
  title = {Dynamics at the air-water interface revealed by evanescent wave light scattering},
  journal = {Eur.\ Phys.\ J. E},
  year = {2009},
  volume = {29},
  pages = {95-105},
  doi = {10.1140/epje/i2009-10455-1},
}


\end{document}